\documentclass[
 aip,
 jcp,
 amsmath,amssymb,
 preprint,
 floatfix
]{revtex4-1}

\usepackage{graphicx}
\usepackage{dcolumn}
\usepackage{bm}

\newcommand{\ket}[1]{\left| #1 \right\rangle}
\newcommand{\bra}[1]{\left\langle #1 \right|}
\newcommand{\braket}[2]{\left\langle #1 \middle| #2 \right\rangle}
\newcommand{\annop}[1]{\hat{a}_{#1}}
\newcommand{\creop}[1]{\hat{a}_{#1}^{\dagger}}
\newcommand{\exop}[2]{\creop{#1}\annop{#2}}
\newcommand{\tse}[1]{\textsuperscript{\emph{#1}}}

\usepackage{xcolor}

\begin{document}

\title{Surface Hopping Dynamics Including Intersystem Crossing using the Algebraic Diagrammatic Construction Method}

\author{Sebastian Mai}
\email{sebastian.mai@univie.ac.at}
\author{Felix Plasser}
\email{felix.plasser@univie.ac.at}
\affiliation{Institute of Theoretical Chemistry, Faculty of Chemistry, University of Vienna, W\"ahringer Str. 17, 1090 Vienna, Austria}
\author{Mathias Pabst}
\affiliation{Institute of Physical Chemistry, University of Mainz, Duesbergweg 10, D-55099 Mainz}
\author{Frank Neese}
\affiliation{Max Planck Institute for Chemical Energy Conversion, Stiftstr. 34-36, D-45470 M\"{u}lheim an der Ruhr, Germany}
\author{Andreas K\"{o}hn}
\affiliation{Institute of Physical Chemistry, University of Mainz, Duesbergweg 10, D-55099 Mainz}
\affiliation{Institute of Theoretical Chemistry, University of Stuttgart, Pfaffenwaldring 55, D-70569 Stuttgart, Germany}
\author{Leticia Gonz\'{a}lez}
\affiliation{Institute of Theoretical Chemistry, Faculty of Chemistry, University of Vienna, W\"ahringer Str. 17, 1090 Vienna, Austria}%

\date{\today}

\begin{abstract}
We report an implementation for employing the algebraic diagrammatic construction to second order [ADC(2)] ab initio electronic structure level of theory in nonadiabatic dynamics simulations in the framework of the SHARC (surface hopping including arbitrary couplings) dynamics method.
The implementation is intended to enable computationally efficient, reliable, and easy-to-use nonadiabatic dynamics simulations of intersystem crossing in organic molecules.
The methodology is evaluated for the 2-thiouracil molecule.
It is shown that ADC(2) yields reliable excited-state energies, wave functions, and spin-orbit coupling terms for this molecule.
Dynamics simulations are compared to previously reported results using high-level multi-state complete active space perturbation theory, showing favorable agreement.
\end{abstract}

\maketitle


\section{Introduction}

Intersystem crossing (ISC) is a fundamental photophysical process which can occur after a molecule is excited by the absorption of light.
Specifically, during ISC the excited-state population is nonradiatively transferred between electronic states of different multiplicity, i.e., a change of the total spin of the electronic wave function occurs.
ISC is usually contrasted with internal conversion (IC), which is a population transfer between states of the same multiplicity.

ISC plays an essential role in photoinduced processes\cite{Klessinger1995,Turro2009} in different areas of research.
For example, in light harvesting,\cite{Segura2005CSR,Colvin2011JACS} OLED design,\cite{Uoyama2012N} magnetic data storage,\cite{Epstein2003MB} and molecular electronics,\cite{Gutlich1994ACIEE,Roy2000APL} ISC is often responsible for improving the efficiency of the application or even for enabling it in the first place.
ISC is also relevant for several biological processes, like oxygen binding to proteins,\cite{Saito2014JPCB} DNA photodamage,\cite{Barbatti2014TCC} or photodynamic therapy.\cite{Dolmans2003NRC}
While traditionally, ISC was regarded as much slower than IC---due to the fact that ISC is completely forbidden in a non-relativistic framework---nowadays it has been shown that ISC also occurs on ultrafast time scales in the pico- to femtosecond range.
This is true not only for diverse transition metal complexes,\cite{Yersin2004TCC,McCusker2003ACR,Forster2006CCR,Vlvcek2010,Cannizzo2010CCR,Aubock2015NC} but also for small (bio-)organic molecules without very heavy atoms, like ketones\cite{Aloise2008JPCAa,Huix-Rotllant2013PCCP} or modified nucleobases.\cite{Kobayashi2009JPCA,Matsika2014TCC,Pollum2014TCC}

The theoretical description of ISC---especially ultrafast ISC---is very challenging.
On the one hand, ISC is a relativistic effect and is mediated by spin-orbit couplings (SOCs).
There are multiple challenges involved when doing accurate relativistic quantum chemistry, for example due to the (bi-)spinor nature of the wave functions, due to the negative-energy continuum of the eigenspectrum of the Dirac/Breit equations, or due to the high 1- and N-particle basis set requirements.\cite{Dyall2007,Reiher2009}
On the other hand, it is often necessary to include the coupling between electronic and nuclear degrees of freedom.
Whereas for slow ISC processes it is possible to use perturbative approaches, e.g., Fermi's Golden Rule,\cite{Marian2012WCMS} explicit dynamics simulations are required to simulate ultrafast ISC.
It is possible to simulate ISC dynamics by means of grid-based quantum dynamics or multi-configurational time-dependent Hartree.\cite{Eng2015ACR, Fumanal2017JCTC,Capano2017PCCP}
However, these methods rely on a careful parameterization of the model used and are often severely hampered by the fact that they can only include a small number of nuclear degrees of freedom.
Therefore, in the last years several groups have established nonadiabatic dynamics methods compatible with the on-the-fly computation of the potential energy surfaces.
The surface hopping method\cite{Tully1990JCP,Barbatti2011WCMS} has been shown to be particularly viable for this purpose\cite{Granucci2012JCP,Cui2014JCP,Mai2015IJQC,FrancodeCarvalho2015JCP} and also the related ab initio multiple spawning\cite{Curchod2016JCP} has been applied.
Over the last years some of us have devoted considerable effort to the \textsc{Sharc} (surface hopping including arbitrary couplings)\cite{Richter2011JCTC,Mai2015IJQC} code, which is a freely available nonadiabatic dynamics package\cite{Mai2014SHARC} that allows performing IC and ISC dynamics with various electronic structure methods implemented in different quantum chemistry packages.

Nonadiabatic dynamics simulations for ISC rely on a proper choice of an electronic structure method, because this choice affects the accuracy of the calculation, possibly more than any other simulation parameter.
The chosen electronic structure method does not only need to provide accurate energies, but it also has to produce energy gradients, nonadiabatic coupling terms, and SOCs to be compatible with surface hopping for ISC.
At the same time, it has to be efficient enough to allow for the $10^4$ to $10^5$ single-point calculations required for a typical ensemble of trajectories.
In the past, a number of different electronic structure methods was employed for such ISC-focused simulations.
Semi-empirical methods offer a computationally efficient possibility\cite{Granucci2012JCP, Martinez-Fernandez2014CS,Favero2016PCCP} although a careful parameterization of the Hamiltonian is required.
Within ab initio methods, one popular choice is the complete active space self-consistent field (CASSCF) method.\cite{Richter2014PCCP,Marazzi2016JPCL,Pederzoli2017,Bellshaw2017CPL}
Unfortunately, CASSCF lacks dynamical correlation.
The application of multi-reference methods with dynamical correlation, like multi-state complete active space perturbation theory (MS-CASPT2)\cite{Tao2009JPCA,Mai2016JPCL,Park2017JCTC} or multi-reference configuration interaction (MRCI),\cite{Mai2014JCP_SO2, Mai2016NC,Borin2017PCCP,Peccati2016PTRSA} significantly improves the accuracy of the simulations, but also dramatically increases their computational expense.
Among the single-reference electronic structure methods, so far only time-dependent density functional theory (TDDFT) has been applied for ISC-focused dynamics simulations.\cite{Atkins2017JPCL, FrancodeCarvalho2015JCP}
Although here computational efficiency and dynamical correlation are in principle given, in TDDFT the choice of the appropriate exchange-correlation functional can strongly affect the quality of the results.

In order to expand the range of applicable electronic structure methods for ISC dynamics simulations, we have extended the nonadiabatic dynamics package \textsc{Sharc}\cite{Richter2011JCTC,Mai2015IJQC,Mai2014SHARC} to use the ab initio algebraic diagrammatic construction at second order (ADC(2)) method from the \textsc{Turbomole} electronic structure suite.\cite{TURBOMOLE70}
Previously, nonadiabatic dynamics simulations with this method were only possible within one multiplicity, i.e., only for IC but not ISC.\cite{Plasser2014JCTC}
The present implementation allows for the computation of nonadiabatic coupling terms within the singlet and triplet manifolds, as well as SOCs between excited singlet and triplet states.
It is thus possible to model IC within the singlet and triplet manifolds and ISC between them.
Currently, the computation of triplet-triplet SOCs is not supported.
Triplet-triplet SOCs (e.g., routinely available in multi-configurational methods) can be of similar magnitude as singlet-triplet SOCs; however, their importance for nonadiabatic dynamics is often minor, as triplet-triplet IC is dominated by nonadiabatic couplings.
Only if the nonadiabatic couplings are very weak (e.g., symmetry forbidden\cite{Mai2014JCP_SO2}) or the SOCs very large (e.g., transition metal complexes\cite{Fumanal2017JCTC}) triplet-triplet SOCs will notably influence the dynamics.
In organic molecules, it can be expected that triplet-triplet SOCs are not of prime relevance.

The idea of computing SOCs with ADC(2) (and CC2) has been initially introduced by Pabst and K\"{o}hn\cite{Pabst2011PhD} in 2011.
This methodology, as implemented in \textsc{Turbomole}, is used here.
More recently, Krauter et al.\cite{Krauter2017CP} and Helmich-Paris et al.\cite{Helmich-Paris2016JCTC} reported on further implementations of SOCs for ADC(2) and CC2, respectively.
The implementation by Krauter et al.\cite{Krauter2017CP} in \textsc{Q-Chem} employs an efficient algorithm for computing exact state-to-state transition moments\cite{Knippenberg2012} using the intermediate-state representation.\cite{Schirmer2004}
In contrast, ADC(2) state-to-state transition moments in the \textsc{Turbomole} implementation are computed approximately by restricting the terms to those appearing in the closely related CC2 response theory.\cite{Pabst:JCP129-214101}
Furthermore, \textsc{Turbomole} features a powerful resolution-of-the-identity (RI) approximation,\cite{Haettig:JCP113-5154} which allows for extremely efficient excited-state computations.
We note here that, in addition to specific SOC implementations, there also exist general SOC implementations, like the recent PySOC,\cite{Gao2017JCTC} which could in principle be employed in nonadiabatic dynamics simulations like the ones targeted here.

\begin{figure}
\includegraphics[scale=1]{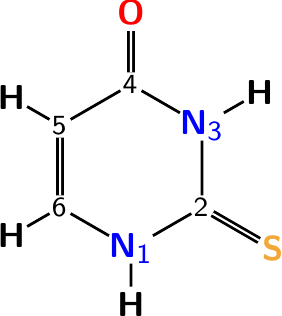}
\caption{Structure of the most stable tautomer of 2-thiouracil (2TU) and atom numbering. }
\label{fig:struc}
\end{figure}

The power of the newly implemented method is demonstrated for the case of 2-thiouracil (2TU, Figure~\ref{fig:struc}), an analogue of uracil where one oxygen atom has been replaced by sulfur.
The excited-state dynamics of 2TU was intensively studied---both experimentally\cite{Pollum2014JCP,Yu2016PCCP,Sanchez-Rodriguez2017PCCP} and theoretically\cite{Cui2013JCP,Gobbo2014CTC,Mai2015JPCA,Mai2016JPCL}---in the last few years, due to its biological relevance and the fact that it exhibits ultrafast ($<$1~ps) ISC with nearly 100\% quantum yield.
Here, we compare excited-state dynamics obtained with the new RI-ADC(2) implementation to previously reported MS-CASPT2-based dynamics simulations.\cite{Mai2016JPCL}
The two methods agree in terms of the overall ISC times, although they differ in some of the mechanistic details.
At the same time, RI-ADC(2) dynamics allows for dramatic computational savings.

\section{Methods}
\label{sec:methods}
\subsection{Ab initio surface hopping dynamics}

The excited-state dynamics simulations with \textsc{Sharc} are performed in a fully diagonalized state basis, as detailed in Ref.~\onlinecite{Mai2015IJQC} and briefly summarized below.
In the case of ISC dynamics, the total electronic Hamiltonian needs to contain the SOCs.
Due to the complexity of full 2- or 4-component electronic structure methods, \textsc{Sharc} is based on a perturbational ansatz, where the Hamiltonian is split into the \textit{molecular Coulomb Hamiltonian} (MCH) and the spin-orbit Hamiltonian:
\begin{equation}
  \hat{H}_\text{total}
  =
  \hat{H}_\text{MCH}
  +
  \hat{H}_\text{SOC}.
\end{equation}
The electronic structure problem is first solved for $\hat{H}_\text{MCH}$, yielding a small set of lowest-energy states (called the MCH states here) of different multiplicities, e.g., singlets and triplets.
In the space of these states, the matrix representation of $\hat{H}^\text{el}_\text{total}$ in the basis of the MCH states (written as $\mathbf{H}^\text{MCH}$) is computed and diagonalized to yield the \textit{diagonal} energies and eigenvectors:
\begin{equation}
  \mathbf{H}^\text{diag}=\mathbf{U}^\dagger\mathbf{H}^\text{MCH}\mathbf{U},
\end{equation}
where $\mathbf{U}$ is the transformation matrix between the MCH and diagonal bases.

The diagonal entries of $\mathbf{H}^\text{diag}$ are the eigenenergies of the total Hamiltonian, which are the energies on which the nuclear dynamics is simulated.
The nuclear motion is governed by the classical equation of motion, which is integrated by the velocity-Verlet algorithm.\cite{Verlet1967PR}
The required forces are calculated by transforming the gradients of the MCH states into the diagonal basis.

As with all surface hopping methods,\cite{Tully1990JCP,Barbatti2011WCMS} the forces are evaluated for the current \textit{active state}.
In order to find the active state, the total electronic wave function is written as a linear combination of the diagonal states:
\begin{equation}
  \ket{\Psi^\text{el}}(t)
  =
  \sum_\alpha
  c_\alpha(t)\ket{\Psi^\text{diag}_\alpha(t)},
\end{equation}
which is propagated from time $t$ to time $t+\Delta t$ by:
\begin{equation}
  \mathbf{c}^\text{diag}(t+\Delta t)
  =
  \mathbf{U}^\dagger(t+\Delta t)
  \mathbf{P}^\text{MCH}(t+\Delta t,t)
  \mathbf{U}(t)
  \mathbf{c}^\text{diag}(t).
  \label{eq:propag}
\end{equation}
The propagator matrix $\mathbf{P}^\text{MCH}(t+\Delta t,t)$ can be computed by appropriately integrating over the Hamiltonian matrix and the nonadiabatic coupling terms from $t$ to $t+\Delta t$.
Instead of using the nonadiabatic couplings, the propagator matrix could also be calculated from the overlap matrix ($S_{IJ}=\braket{\Psi_I(t)}{\Psi_J(t+\Delta t)}$) using the \textit{local diabatization} method.\cite{Granucci2001JCP, Plasser2012JCP}
From the change of the wave function coefficients $\mathbf{c}^\text{diag}$, the active state of the next time step is found stochastically.\cite{Mai2015IJQC}

From the above-said, it follows that several properties need to be calculated at each simulation time step:
(i) energies, (ii) gradients, (iii) SOCs, and (iv) nonadiabatic couplings \textit{or} overlap matrices.
For setting up initial conditions (or explicit interactions with a laser field) also dipole and transition dipole moments are needed.
Any quantum chemistry method/program used together with \textsc{Sharc} needs to provide these quantities, in an efficient and consistent manner.


\subsection{ADC(2) excitation energies and transition densities}

In ADC(2) theory,\cite{Schirmer:PRA26-2395,Trofimov1995JPBAMOP} excitation energies are obtained from an eigenvalue equation of the form
\begin{equation}
 \label{eq:adc2-eve}
 \mathbf A \mathbf R^I = \mathbf R^I \omega_I 
\end{equation}
where the eigenvectors consist of a single excitation part $\mathbf R_1$ (associated with single excitation operators $\hat\tau_{\rho_1}$ relative to the Hartree-Fock reference state $\ket{\Phi_0}$) and a double excitation part $\mathbf R_2$ (double excitation operators $\hat\tau_{\rho_2}$). 
Here, we use the general indices $\rho_1$, $\sigma_1$, and $\rho_2$, $\sigma_2$ to refer to single and double excitation manifolds, respectively.
An explicit representation for singlet and triplet excitations is given in the Appendix \ref{app:ttd}.
The symmetric matrix $\mathbf A$ can be written in blocked form as
\begin{equation}
 \label{eq:adc2-defA}
 \mathbf A = \begin{pmatrix} A_{\rho_1\sigma_1} &  A_{\rho_1\sigma_2}\\
                             A_{\rho_2\sigma_1} &  A_{\rho_2\sigma_2} \end{pmatrix}
\end{equation} 
with
\begin{eqnarray}
 \label{eq:adc2-defAmatels}
  A_{\rho_1\sigma_1} &=&\hat{\mathcal{S}}_{\rho_1\sigma_1}\bra{\Phi_0}\hat\tau_{\rho_1}^\dagger[\hat H + [\hat H,\hat T_2],\hat\tau_{\sigma_1}] \ket{\Phi_0} \\
  A_{\rho_1\sigma_2} &=&\bra{\Phi_0}\hat\tau_{\rho_1}^\dagger[\hat H,\hat\tau_{\sigma_2}]\ket{\Phi_0} \\
  A_{\rho_2\sigma_1} &=& \bra{\Phi_0}\hat\tau_{\rho_2}^\dagger[\hat H,\hat\tau_{\sigma_1}]\ket{\Phi_0} \\    
  A_{\rho_2\sigma_2} &=&\bra{\Phi_0}\hat\tau_{\rho_2}^\dagger[\hat F,\hat\tau_{\sigma_2}]\ket{\Phi_0} 
\end{eqnarray} 
Here, $\hat H$ is the molecular electronic Hamiltonian and $\hat F$ the Fock operator corresponding to the reference determinant  $\ket{\Phi_0}$. 
$\hat T_2$ is the pair correlation cluster operator, determined from M{\o}ller-Plesset second-order perturbation (MP2) theory and $\hat{\mathcal{S}}_{\rho_1\sigma_1}$ is a symmetrization operator: $\hat{\mathcal{S}}_{\rho_1\sigma_1}  A_{\rho_1\sigma_1} =  \frac{1}{2}\left(A_{\rho_1\sigma_1} + A_{\sigma_1\rho_1} \right)$ . 
In order to arrive at potential energy surfaces of excited states, the total energy of a state $I$ is defined as $E_I = E_0^{\text{MP2}}+\omega_I$.
In the course of this work, we view ADC(2) as a simplified variant of the CC2 method.\cite{Haettig:AQC50-37} The transition between the methods is easily accomplished by replacing the CC2 ground state equations by MP2 (effectively setting the $\hat T_1$ cluster operator to zero) and symmetrization of the matrix $\mathbf A$. 
For details of the implementation, we refer to the original works on CC2 using the resolution-of-the-identity (RI) approximation for fast evaluation of the two-electron repulsion integrals.\cite{Haettig:JCP113-5154} 
The particular feature of this implementation is a recasting of the eigenvalue problem (Eq. \ref{eq:adc2-eve}) into an effective (energy-dependent) eigenvalue problem for single excitations $\mathbf R_1$ only. 
The double excitation contribution can always be computed from the known  $\mathbf R_1$ amplitudes, thus $O(N^4)$ scaling storage of doubles amplitudes is fully avoided.\cite{Haettig:JCP113-5154}

In the course of the present work, we need to define transition matrix elements between electronic ground or excited singlet states and triplet states. 
To this end, we again start from the usual response theory for coupled-cluster theory within the CC2 context and derive the ADC(2) expressions using the simplifications as explained in the previous paragraph. 
We note that this approach deviates from the true ADC(2) philosophy, which would require parts of the second-order ground state wave function in order to arrive at matrix elements that are fully correct through second-order perturbation theory.\cite{Krauter2017CP}
For the definition and the general outline of computing transition densities between the electronic ground state and excited states and between excited states within the CC2 context, we refer to Refs. \onlinecite{Haettig:JCP117-6939} and \onlinecite{Pabst:JCP129-214101}. 
In the spin-orbital formulation, these expressions are easily generalized to triplet transition densities, i.e., transition densities of the spin-excitation operator $\hat{\mathbf{E}}_{pq}^{\text{spin}}$ whose Cartesian components are given as\cite{Berning2000MP}
\begin{eqnarray}
\label{eq:spinop1}
\hat{E}_{pq}^{x}=\frac{1}{2}\left(\exop{p}{\bar q}+\exop{\bar p}{q}\right)\\
\label{eq:spinop2}
\hat{E}_{pq}^{y}=\frac{\mathrm{i}}{2}\left(\exop{\bar p}{q}-\exop{p}{\bar q}\right)\\
\label{eq:spinop3}
\hat{E}_{pq}^{z}=\frac{1}{2}\left(\exop{p}{q}-\exop{\bar p}{\bar q}\right)
\end{eqnarray}
Here, $\creop{p}$ and $\annop{p}$ are the usual creation and annihilation operators and the bar is used to indicate the orbitals of $\beta$ spin. 
Here and in the following, $p$ and $q$ are the indices of general (occupied and virtual) spatial orbitals.
As the implementation treats only the $M_S=0$ component of the triplet state explicitly, only transition moments of the $\hat{E}_{pq}^{z}$ component can be directly computed. The other components, however, are easily obtained by the Wigner-Eckart theorem (for details see, e.g., Ref.\ \onlinecite{Krauter2017CP}).

A special feature of the present implementation is the use of a spin-adapted basis for the triplet operators, allowing to use the fast spin-adapted code for the ground state of closed-shell systems. The corresponding expressions for triplet transition densities are given in Appendix \ref{app:ttd}.

\subsection{Spin-orbit couplings}
The SOC terms are defined here as the matrix elements of the Breit-Pauli spin-orbit Hamiltonian\cite{Reiher2009}
\begin{equation}
\label{eq:SOBP}
\widehat{H}_\text{SO,BP}=
\frac{1}{2c^2}\sum_{i=1}^{n_{el}}\sum_{K=1}^{n_{nuc}}
\frac{Z_K(\mathbf{r}_{iK}\times \mathbf{p}_i)\cdot \mathbf{s}_i}{r_{iK}^3}
-\frac{1}{2c^2}\sum_{i,j\neq i}^{n_{el}}\frac{(\mathbf{r}_{ij}\times \mathbf{p}_i)\cdot \mathbf{s}_i}{r_{ij}^3}+
\frac{1}{c^2}\sum_{i,j\neq i}^{n_{el}}\frac{(\mathbf{r}_{ij}\times \mathbf{p}_i)\cdot \mathbf{s}_j}{r_{ij}^3}
,
\end{equation}
which is given with respect to the Cartesian position $\mathbf{r}_{iK},\mathbf{r}_{ij}$, momentum $\mathbf{p}_i$, and spin $\mathbf{s}_i$ operators of the different electrons.
The terms in Eq.~\eqref{eq:SOBP} are a one-electron term and the two-electron spin-same-orbit and spin-other-orbit interaction contributions.
From a practical viewpoint it is worth noting that the effort for computing a matrix element of $\widehat{H}_{\text{SO},\text{BP}}$ exceeds that of the spin-free Hamiltonian due to the fact that more two-electron integrals have to be evaluated.
However, the computational effort can be reduced by applying a mean field approach.\cite{Berning2000MP, Neese2005JCP}
For this purpose an effective spin-orbit mean field (SOMF) operator is constructed, which is of the form
\begin{equation}
\widehat{H}_{\text{SOMF}}=
\sum_{pq}\mathbf{f}_{pq}\cdot \hat{\mathbf{E}}_{pq}^{\text{spin}}=
\sum_{\zeta\in \lbrace x,y,z\rbrace}\sum_{pq}f_{pq}^{\zeta}\hat{E}_{pq}^{\zeta}
\end{equation}
with the spin-excitation operator $\hat{\mathbf{E}}_{pq}^{\text{spin}}$ as defined above in Eqs. \eqref{eq:spinop1} to \eqref{eq:spinop3}.

The matrix elements $f_{pq}^{\zeta}$ are obtained by averaging the two-electron contributions over the Hartree-Fock one-particle density, in analogy to the definition of the Fock matrix\cite{Berning2000MP}
\begin{equation}
\label{eq:somf}
f_{pq}^{\zeta}=
h^{\text{SOC},\zeta}_{pq}+
\sum_{k}\left(
2 g^{\text{SOC},\zeta}_{kkpq}-3 g^{\text{SOC},\zeta}_{pkkq}+ 3 g^{\text{SOC},\zeta}_{qkkp}
\right)
\end{equation}
Here, $h^{\text{SOC}}_{pq}$, $g^{\text{SOC},\zeta}_{pqrs}$ are the required one- and two-electron SOC integrals.
The SOC between two wave functions $\Psi_I$ and $\Psi_J$ is in turn obtained through contraction with the spin-transition density matrix \cite{Berning2000MP}
\begin{eqnarray}
\label{eq:cont_somf}
\bra{\Psi_I}\widehat{H}_{\text{SOMF}}\ket{\Psi_J}&=&
\sum_{\zeta}\sum_{pq}D_{pq}^{IJ,\zeta}f_{pq}^{\zeta}\\
D_{pq}^{IJ,\zeta}&=&\bra{\Psi_I}\hat{E}_{pq}^{\zeta}\ket{\Psi_J}
\end{eqnarray}

\subsection{Nonadiabatic interactions}
Nonadiabatic interactions are computed during the dynamics by a formalism initially proposed by Hammes-Schiffer and Tully,\cite{Hammes-Schiffer1994JCP} and later refined by Persico and coworkers.\cite{Granucci2001JCP}
In this approach, the inner product of the nonadiabatic coupling vector and the velocity, which determines the nonadiabatic transition probabilities, is not computed explicitly.
Instead, the wave function overlap
\begin{equation}
\label{eq:SIJ}
S_{IJ}(t)=
\braket{\Psi_I(t)}{\Psi_J(t+\Delta t)}
\end{equation}
between the electronic wave functions computed at two different time steps is used in the computation of the propagator matrix in equation~\eqref{eq:propag}.

For computational efficiency and ease of implementation, the ADC(2) wave functions are approximated to be of the form of configuration interaction singles (CIS) wave functions using
\begin{equation}
\ket{\Psi_I} = \sum_{ia}\frac{R_{ia}^{I}}{||\mathbf{R}_1^{I}||}\ket{\Phi_{ia}}
\end{equation}
where $R_{ia}^{I}$ is the amplitude of the single excitation going from occupied orbital $i$ to virtual orbital $a$ and $\ket{\Phi_{ia}}$ is the corresponding Slater determinant.
The idea of using approximate CIS wave functions for nonadiabatic dynamics has been previously used very successfully in the case of TDDFT\cite{Tapavicza2007PRL, Mitric2008JCP, Pittner2009CP} and the formalism has been extended to the case of ADC(2) more recently.\cite{Plasser2014JCTC}
The evaluation of Eq.~\eqref{eq:SIJ} can be very costly even for the approximate wave functions.
To allow for an efficient computation of this term, we adapted the optimized algorithm described in Ref.~\citenum{Plasser2016JCTC}.


\section{Computational Details}

The computations are divided into two main parts.
First, we present benchmark calculations, where we compute a few excited states at the Franck-Condon geometry of 2TU with RI-ADC(2),\cite{Trofimov1995JPBAMOP, Haettig:AQC50-37} ADC(3),\cite{Harbach2014, Dreuw2015} and MS-CASPT2,\cite{Andersson1990JPC} and use the two latter methods to scrutinize the accuracy of RI-ADC(2) for this molecule.
These calculations were performed at the RI-MP2/def2-SVP optimized ground state minimum (see the supporting information (SI) for coordinates).
In the second part, we perform nonadiabatic dynamics simulations with \textsc{Sharc} coupled to RI-ADC(2).

The RI-ADC(2)\cite{Trofimov1995JPBAMOP, Haettig:AQC50-37} computations were performed using \textsc{Turbomole} 7.0\cite{TURBOMOLE70} and employed the def2-SVP basis set.\cite{Weigend2005PCCP}
For comparison, also a computation using the larger aug-cc-pVTZ basis set\cite{Dunning1989JCP} was performed.
The spin-orbit matrix elements were computed using the spin-orbit mean field (SOMF) formalism as implemented in the REL module of the \textsc{Orca} 3.0.3 program package.\cite{Neese2005JCP, Neese2012WCMS}
The code was also tested for \textsc{Orca} version 4.0.1.
We use the seminumerical implementation of the Coulomb-like contributions to the spin-orbit mean-field integrals, while the exchange-like contributions are approximated by a one-center approximation.\cite{Neese2005JCP}
We note that future releases of \textsc{Turbomole} will also directly provide spin-orbit mean-field integrals, computed without further approximations, see Ref.\ \onlinecite{Helmich-Paris2016JCTC}.

ADC(3) computations,\cite{Harbach2014, Dreuw2015} also employing the def2-SVP basis, were carried out with the \textsc{Q-Chem} program package.\cite{Shao2006}
In this case, a canonical implementation without RI approximation was used.
Density matrices at the ADC(3) level were obtained by contracting the third order vectors with the second order intermediate-state representation.\cite{Schirmer2004}

MS-CASPT2\cite{Andersson1990JPC} computations were performed with \textsc{Molcas} 8.0\cite{Aquilante2015JCC} using an active space of 14 electron in 10 orbitals ($n_S, n_O, 5\times\pi, 3\times\pi^*$; see SI for orbital plots) considering 6 singlet and 4 triplet states, i.e., MS(6,4)-CASPT2(14,10).
The orbitals for the present computation were obtained by a state-averaged complete active space self-consistent field (SA-CASSCF)\cite{Roos1980CP} computation using the same active space and state-averaging.
For these computations, the SVP basis set of \textsc{Molcas} was used, which is identical to def2-SVP in the case of 2TU.
Following Ref.~\citenum{Zobel2017CS}, for this rather small basis set we use an IPEA shift of zero,\cite{Ghigo2004CPL} which produces energies in good agreement with MS-CASPT2 calculations using much larger basis sets (up to quadruple-$\zeta$).\cite{Mai2015JPCA}
Spin-orbit matrix elements were computed using the state interaction method combined with an atomic mean-field Hamiltonian.\cite{Malmqvist2002CPL,Hess1996CPL}
Note that our previous calculations for 2TU\cite{Mai2016JPCL} were performed with MS(3,3)-CASPT2(12,9).

The produced wave functions were analyzed by visualizing natural transition orbitals.\cite{Martin2003} 
To this end, the TheoDORE~1.5 program package\cite{Plasser2012JCTC, Plasser2014JCP1, Plasser2017TheoDORE} was used for the analysis of the \textsc{Turbomole} computations, whereas integrated analysis modules were employed for the results of the the \textsc{Q-Chem}\cite{Plasser2014JCP1} and \textsc{Molcas}\cite{Plasser_sub_Molcas} program packages.

Dynamics simulations were performed at the same RI-ADC(2)/def2-SVP level of theory as the benchmark excited-state calculations.
Nonadiabatic interaction terms were computed by means of wave function overlaps,\cite{Plasser2016JCTC} which were used in the local diabatization procedure to propagate the wave function.\cite{Granucci2001JCP, Plasser2012JCP}
In preparation of the dynamics simulations, 200 initial geometries were sampled from a Wigner distribution of the ground state harmonic oscillator, based on a frequency calculation at the MP2/cc-pVDZ level of theory (as in Refs.~\onlinecite{Mai2016JPCL}).
For each geometry, a single point calculation at the RI-ADC(2)/def2-SVP level of theory was performed, and the resulting excitation energies and oscillator strengths were used to simulate the absorption spectrum of 2TU (see below).
The energies and oscillator strengths were employed to stochastically select bright initial states\cite{Barbatti2007JPPA} in the 3.9--4.3~eV energy window, which yielded 32 initial conditions for trajectories (18 starting in $S_2$ and 14 in $S_3$).
The trajectories were propagated with \textsc{Sharc},\cite{Richter2011JCTC,Mai2015IJQC,Mai2014SHARC} considering 4~singlet and 3~triplet states, for 1000~fs with a 0.5~fs nuclear time step and a 0.02~fs electronic time step.
An energy-based decoherence scheme was applied to the diagonal states.\cite{Granucci2007JCP}
For the analysis, out of 32~trajectories, 3 were neglected because they showed a ring opening of 2TU, and RI-ADC(2) is not expected to provide reliable results in this situation due to the small $S_1-S_0$ energy gap.
The results of the analysis of the 29~remaining trajectories are presented below.


\section{Results and Discussion}

\subsection{Vertical excitations and SOC terms}

The purpose of this section is to evaluate the overall ability of the RI-ADC(2) method to compute excited state energies and wave functions at the Franck-Condon region, and to evaluate specifically the SOC terms.
In Table~\ref{tab:en} the main results of the RI-ADC(2)/def2-SVP computations at the ground-state minimum are presented and compared to two higher-level reference methods: ADC(3)/def2-SVP and MS(6,4)-CASPT2(14,10)/def2-SVP.
For RI-ADC(2) also the results using the larger aug-cc-pVTZ basis set are given (in parentheses).
For the double-$\zeta$ calculations, the corresponding natural transition orbitals\cite{Martin2003, Plasser2014JCP1} are given in the supplemental material.

\begin{table*}
\caption{Vertical excitation energies ($\Delta E$, eV), oscillator strengths ($f$), and dipole moments ($\mu$, D) at the MP2-optimized ground-state minimum of 2TU computed at the RI-ADC(2)/def2-SVP, ADC(3)/def2-SVP and MS(6,4)-CASPT2(14,10)/def2-SVP levels of theory.}
\label{tab:en}
\begin{tabular}{l|ccc|ccc|ccc}
\hline
State\tse{a} & \multicolumn{3}{c|}{RI-ADC(2)} &  \multicolumn{3}{c|}{ADC(3)} &  \multicolumn{3}{c}{MS-CASPT2} \\ 
 & $\Delta E$\tse{b}  & $f$ & $\mu$
 & $\Delta E$  & $f$ & $\mu$
 & $\Delta E$  & $f$ & $\mu$
\\
 \hline
$^1 n_S\pi_2^*$ & 3.90 (3.75) & 0.000 & 4.5 & 4.15 & 0.000 & 4.9 & 3.81 & 0.000 & 5.1 \\ 
$^1 n_O\pi_6^*$ & 4.74 (4.58) & 0.000 & 1.5 & 5.19 & 0.000 & 1.7 & 4.64 & 0.000 & 2.2 \\ 
$1^1\pi_S\pi^*$ & 4.78 (4.42) & 0.377 & 4.5 & 4.86 & 0.253 & 5.6 & 4.32 & 0.421 & 5.4 \\ 
$2^1\pi_S\pi^*$ & 5.28 (4.92) & 0.107 & 4.6 & 5.33 & 0.227 & 4.7 & 4.99 & 0.353 & 4.7 \\ 
$^1 n_S\pi_6^*$ & 5.58 (5.28) & 0.000 & 7.6 & 5.98 & 0.000 & 9.0 & 5.19 & 0.001 & 4.0 \\ 
\hline
$^3\pi_S\pi_2^*$ & 3.50 (3.42) &  & 3.99 & 3.37 &  & 3.77 & 3.34 &  & 3.80 \\ 
$^3n_S\pi_2^*$ & 3.75 (3.64)  &  & 4.70 & 4.00 &  & 4.93 & 3.85 &  & 4.56 \\ 
$^3\pi\pi^*$ & 4.07 (3.98) &  & 3.42 & 3.86 &  & 3.65 & 3.85 &  & 3.31 \\ 
$^3n_O\pi_6^*$ & 4.51 (4.40) &  & 1.92 & 4.95 &  & 2.16 & 4.69 &  & 3.03 \\ 
\hline
\end{tabular}

\vspace{0.1em}
\tse{a} The orbital nomenclature is given in supplemental Figure~S1.

\tse{b} RI-ADC(2)/aug-cc-pVTZ results are given in parentheses.
\end{table*}

With all methods, the $S_1$ state is an excitation from the $n$ orbital on the S atom (denoted $n_S$, see Ref.~\citenum{Mai2015JPCA} for a more detailed discussion of the involved orbitals) to the $\pi^*$ orbital located between the S atom and C2 atom ($\pi_2^*$).
This state is located slightly below 4 eV for RI-ADC(2) and MS-CASPT2, and slightly above this value for ADC(3).
The next singlet excitation at the RI-ADC(2) level, denoted $^1 n_O\pi_6^*$, originates from the $n$ orbital on the O atom and goes into a $\pi^*$ orbital located on the C4, C5, and C6 atoms (cf. Figure~\ref{fig:struc}).
Its energy is around 4.7~eV at the RI-ADC(2) and MS-CASPT2 levels whereas it is significantly higher (5.19~eV) at the ADC(3) level.
For all three methods, this state is distinguished by a particularly small dipole moment, around 2~D.
Two bright $\pi\pi^*$ states follow, where in both cases, the donor orbital is a $\pi$ orbital predominantly located on the S atom ($\pi_S$) while the acceptor is a $\pi^*$ orbital delocalized over the whole system, see supplemental Figures S4 and S5 (hence, we simply write ``$\pi^*$'' in the table, without orbital index).
The two resulting $^1\pi_S\pi^*$ states are located at 4.8~eV and 5.3~eV for both the RI-ADC(2) and ADC(3) methods when using the def2-SVP basis set, while they are significantly lower in energy in the case of MS-CASPT2 (4.32 and 4.99 eV).
Interestingly, when the aug-cc-pVTZ basis set is employed, the RI-ADC(2) energies are significantly lower and almost coincide with the MS-CASPT2/def2-SVP energies.
The two $^1\pi_S\pi^*$ states are the only states considered here with significant oscillator strengths.
All methods agree that the lower $^1\pi\pi^*$ state has a slightly enhanced oscillator strength compared to the higher energy one.
The highest considered state ($^1 n_S\pi_6^*$) is an excitation from the S atom to a $\pi^*$ orbital on the opposite side of the ring.
At the RI-ADC(2) and ADC(3) levels, this last state possesses enhanced charge-transfer character as indicated by the large dipole moment (above 7.5~D).
By contrast, the MS-CASPT2 result shows some admixture with the $^1 n_O\pi_6^*$ state (cf. Figure S6), which leads to a signficantly decreased total dipole moment.
It should also be noted that this state possesses substantial contributions of two-electron-excitation character.
Considering both MS-CASPT2 and ADC(3), it is found that this state possesses only 72\% of single-excitation character, as measured by the norm of the one-electron transition density matrix.\cite{Plasser2014JCP1}

In the case of the triplet states the agreement between the three different methods is better.
For the first three triplet states no deviations above 0.25~eV are found among the three methods.
The lowest triplet state, located around 3.4~eV, is of $^3\pi_S\pi^*$ character and shows similar orbital contributions as the two bright singlet states.
Then, two almost degenerate states follow slightly below 4~eV: a $^3n_S\pi_2^*$ state of similar character as the $S_1$ state and a $^3\pi\pi^*$ state, delocalized over the whole system, that possesses no direct counterpart among the computed singlet states.
In the case of the $T_4$ ($^3n_O\pi_6^*$) the discrepancies are somewhat bigger, as RI-ADC(2) and MS-CASPT2 place this state at $\approx$4.6~eV, while it is at 4.95~eV in the case of ADC(3).
This observation is consistent with the energies of the corresponding singlet state ($^1n_O\pi^*_6$).

Table~\ref{tab:en} shows large discrepancies in the electronic excitations already at the Franck-Condon geometry, even for sophisticated methods as ADC(3) and MS-CASPT2.
For the case of 2TU, it is not \textit{a priori} clear which of these two methods is more reliable, i.e., whether it is more important to move to third order in many-body perturbation theory as in the case of ADC(3) or to include multireference effects as in MS-CASPT2.
Moreover, judging the basis set effects is not trivial, considering that MS-CASPT2 with a small basis set and an IPEA shift of zero profits from error compensation,\cite{Zobel2017CS} and hence yields very similar results as MS-CASPT2 with a large CAS(16,12) active space, a quadruple-$\zeta$ basis, and the standard IPEA shift.\cite{Mai2015JPCA}
Therefore, the results have to be carefully checked, individually.
On one hand, it is important to realize that all methods agree satisfactorily in the nature and energy of the $S_1$ state as well as the lowest three triplet states.
Thus, it can be assumed that the dynamics among these methods will be similar once the $S_1$ state is reached and, in particular, that ISC from this state will be described correctly.
On the other hand, the energy of the lowest bright state ($1^1\pi_S\pi^*$) differs significantly between the ADC methods and MS-CASPT2.
In this case, two observations can be made.
First, the ADC(2) energy moves closer to the MS-CASPT2 value once a larger basis set is used.
Second, the MS-CASPT2 energy is actually too low when compared to the experimental absorption spectrum\cite{Khvorostov2005JPCA} (see below).
Finally, the possible influence of the $^1 n_O\pi_6^*$ state on the dynamics has to be examined.
This state is placed at a similar energy as the bright state for RI-ADC(2) while it is significantly higher for the other two methods.
While this is a potential shortcoming of the RI-ADC(2) method, we notice that this state quickly increases in energy during the dynamics and, thus, does not play a role.
We thus summarize that RI-ADC(2)/def2-SVP provides a satisfactory description of all relevant state at the Franck-Condon geometry when compared against the reference methods and experiment.

As a next step, we compare the SOC values between the different states as computed with RI-ADC(2) and MS-CASPT2.
In this case, it should be remembered that a quantitative agreement cannot be expected due to the fact that the wave functions produced by the two methods are not equivalent, as already seen in the $f$ and $\mu$ values shown in Table~\ref{tab:en}.
Thus, only a semi-quantitative agreement can be expected.
In Table~\ref{tab:SOC}, the SOC values are collected for the different pairs of states.
Both methods agree that the largest SOC value ($\approx 125\;\mathrm{cm}^{-1}$) is obtained between the $S_1(^1 n_S\pi_2^*)$ and $T_1(^3\pi_S\pi^*)$ states.
The second largest value ($\approx 100\;\mathrm{cm}^{-1}$) is obtained for the $1^1\pi_S\pi^*/^3n_S\pi_2^*$ state pair.
Then, the $2^1\pi_S\pi^*/^3n_S\pi_2^*$ and $^1 n_S\pi_2^*/^3\pi\pi^*$ follow, both around 75~cm$^{-1}$.
All these values agree reasonably well between the two methods and fall in line with qualitative expectations, i.e., couplings are large between states of different ($n\pi^*$, $\pi\pi^*$) character and if there are strong contributions on the S atom.

\begin{table}
\caption{Comparison of the SOC terms (cm$^{-1}$) between different singlet and triplet excited states computed at the RI-ADC(2)/def2-SVP and MS(6,4)-CASPT2(14,10)/def2-SVP levels of theory for the MP2-optimized ground-state minimum.\tse{a}}
\label{tab:SOC}
\begin{tabular}{l|cccc}
\hline\\*[-0.9em]
 & $^3\pi_S\pi^*$ & $^3n_S\pi_2^*$ & $^3\pi\pi^*$ & $^3n_O\pi_6^*$ \\ 
\hline\\*[-0.9em]
$^1 n_S\pi_2^*$ & \textbf{117/131} & 0/4 & \textbf{68/77} & 4/1 \\ 
$^1 n_O\pi_6^*$ & 3/10 & 4/2 & 32/55 & 1/3 \\ 
$1^1\pi_S\pi^*$ & 0/10 & \textbf{109/83} & 0/4 & 9/9 \\ 
$2^1\pi_S\pi^*$ & 0/0 & \textbf{73/76} & 0/4 & 30/17 \\ 
$^1 n_S\pi_6^*$ & 19/10 & 6/4 & 35/40 & 15/5 \\ 
\hline
\end{tabular}

\tse{a} Results are given in the order RI-ADC(2)/MS-CASPT2. Values above 60 cm$^{-1}$ are marked bold.
\end{table}

Recently, Krauter et al.\cite{Krauter2017CP} have evaluated their ADC/SOC code, implemented in \textsc{Q-Chem}, against the \textsc{Turbomole} interface used here, finding good agreement for the two molecules investigated (thiophene and 1,2-dithiin).
Considering both the results by these authors and the ones in Table~\ref{tab:SOC}, we conclude that the present SOC implementation for RI-ADC(2) produces accurate results.


\subsection{SHARC dynamics}

In this section, we show that the described RI-ADC(2) method, including the calculation of SOCs and wave function overlaps, can be employed to efficiently carry out accurate \textsc{Sharc} dynamics simulations.
To this end, we simulate the excited-state dynamics of 2TU, after excitation to the lowest bright singlet state ($S_2$, with $\pi_S\pi^*$ character in the Franck-Condon region).

\begin{figure*}[t]
  \centering
  \includegraphics[scale=1]{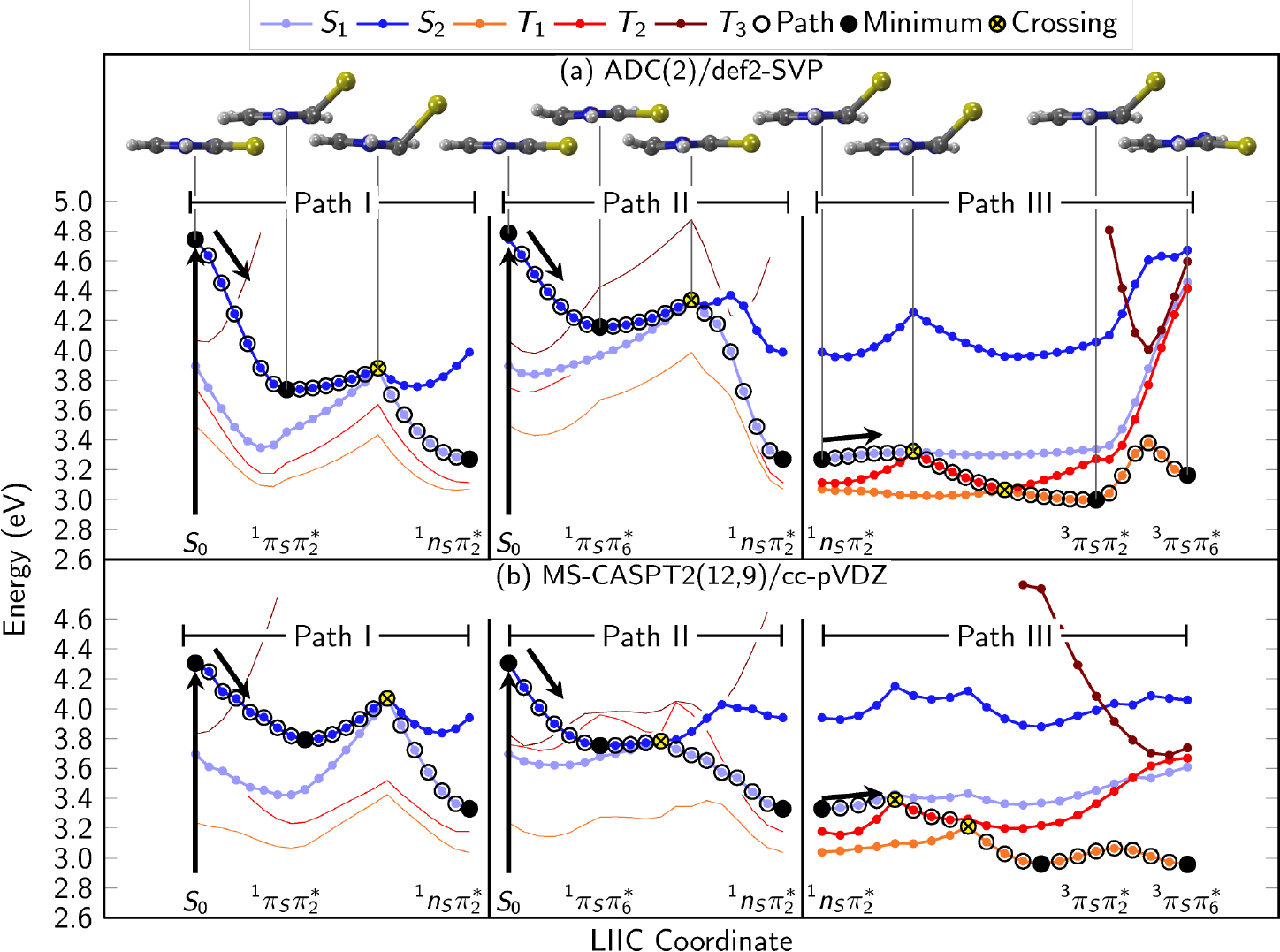}
  \caption{Linear interpolation in internal coordinates (LIIC) scan for the excited states of 2TU at the (a) RI-ADC(2)/def2-SVP and (b) MS-CASPT2(12,9)/cc-pVDZ levels of theory.
  Panel (b) was adapted from S. Mai, P. Marquetand, L. Gonz\'alez, J. Phys. Chem. A 119, 9524 (2015).
  }
  \label{fig:liic}
\end{figure*}

The essential parts of the excited-state PESs of 2TU are presented in Figure~\ref{fig:liic}.
In the upper panel of the figure, we present a chain of linear interpolation in internal coordinates (LIIC) scans between the most important critical points, which were individually optimized with each method.
These scans can be partitioned into three main relaxation paths.
In Path I, after excitation to the $S_2$ state, the molecule first relaxes to one of its two $S_2$ minima, with state character $\pi_S\pi^*_2$ and with a strongly pyramidalized geometry (see the insets).
From this minimum, only a small barrier needs to be overcome to reach a $S_1/S_2$ minimum energy conical intersection (MECI), allowing IC to the $S_1$ ($^1n_S\pi^*_2$ character) and relaxation to the $S_1$ minimum (pyramidalized geometry).
In Path II, from the Franck-Condon geometry, the molecule relaxes to the second $S_2$ minimum ($^1\pi_S\pi^*_6$ character and nearly planar geometry), and from there can reach another $S_1/S_2$ MECI, which again allows IC to the $S_1$.
Path III, which begins in the $S_1$ minimum reached by either Path I or II, leads to an easily accessible $S_1/T_2$ minimum energy crossing point (MECP), where sizable SOCs allow for efficient ISC.
Once in the $T_2$ triplet state, the molecule can relax to one of the two $T_1$ minima via a $T_1/T_2$ MECI.
Finally, a second $T_1$ minimum can be reached by surpassing an energy barrier.

In the lower panel of Figure~\ref{fig:liic}, we also show an analogous scan obtained with MS-CASPT2(12,9)/cc-pVDZ, which is the level of theory employed previously for the study of 2TU.\cite{Mai2015JPCA,Mai2016JPCL}
The comparison between RI-ADC(2) and MS-CASPT2 reveals that both methods yield qualitatively the same three relaxation pathways, as described above.
There are some differences between the methods, most importantly that RI-ADC(2) seems to make passage through Path I easier and through Path II more difficult, compared to MS-CASPT2.
Also, the barrier between the $T_1$ minima is larger with RI-ADC(2) than with MS-CASPT2.
However, in general the agreement is very good.
In particular, both methods predict that after excitation the molecule can easily relax from the initial $S_2$ state to the $S_1$, from where efficient ISC to the $T_2$, followed by relaxation in the triplet states, can commence.

\begin{figure}
  \centering
  \includegraphics[scale=1]{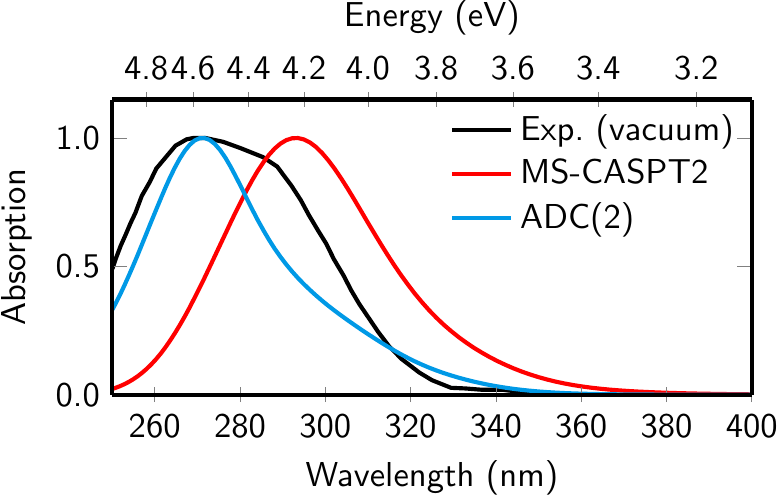}
  \caption{The experimental vacuum absorption spectrum of 2TU\cite{Khvorostov2005JPCA} together with the absorption spectra simulated with MS-CASPT2\cite{Mai2016JPCL} and RI-ADC(2).
  All spectra are normalized.}
  \label{fig:spectrum}
\end{figure}

Encouraged by the good agreement of the MS-CASPT2 and RI-ADC(2) PESs, we went on to perform the \textsc{Sharc} dynamics simulations.
The absorption spectrum, which was generated during the initial condition setup, is shown in Figure~\ref{fig:spectrum}.
The good agreement with the experimental spectrum is another promising indicator of the adequacy of RI-ADC(2) to describe the excited-state dynamicss of this molecule.

\begin{figure}
  \centering
  \includegraphics[scale=1]{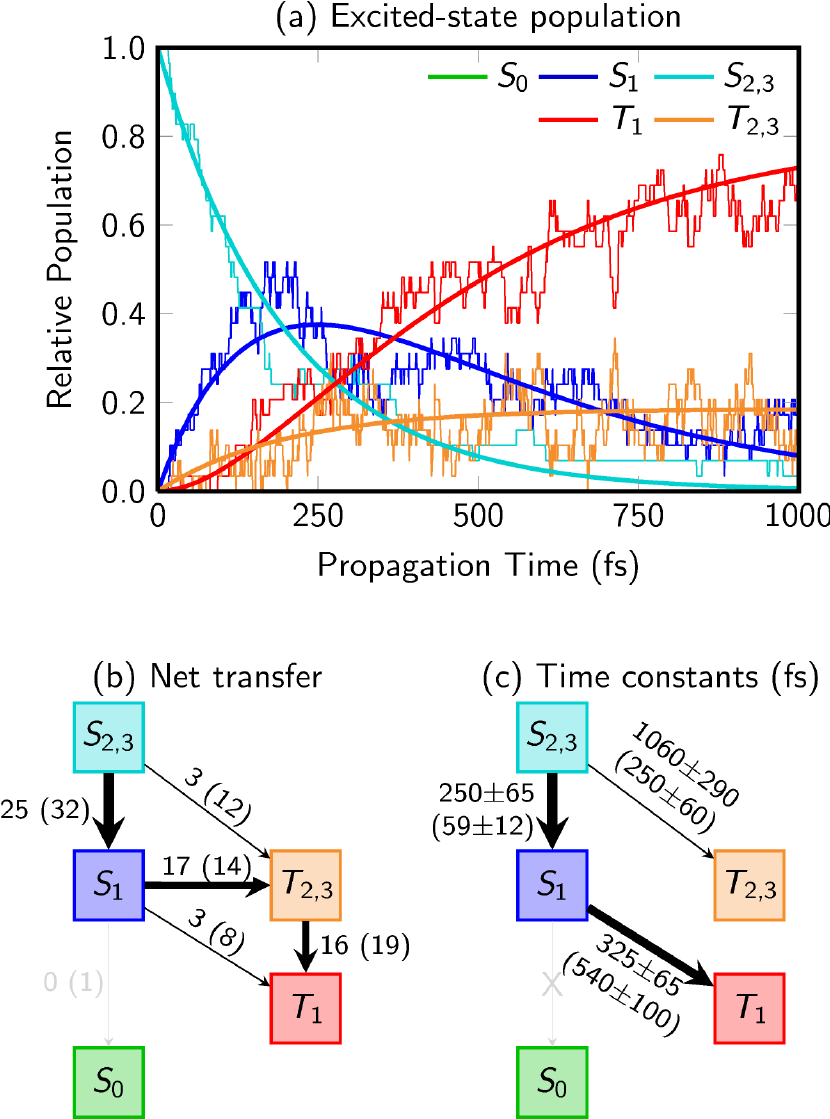}
  \caption{Excited-state populations, net transfer graph, and fitted time constants for the excited-state dynamics simulations of 2TU with RI-ADC(2)/def2-SVP.
  Note that the populations of $S_2$ and $S_3$ as well as $T_2$ and $T_3$ were combined in the figure.
  The analysis includes 29 trajectories.
  In parenthesis, the corresponding values from Ref.~\onlinecite{Mai2016JPCL} are given.}
  \label{fig:pop_const}
\end{figure}

Figure~\ref{fig:pop_const} shows an analysis of the population flow from the RI-ADC(2) dynamics simulations.
Since in the dynamics, it is not unambiguously possible to identify the diabatic state character (e.g., $^1n_S\pi^*_2$, ...), the following paragraphs discuss the dynamics in terms of the eigenstates of the spin-free Hamiltonian (adiabatic, but spin-diabatic)---note that there is not a one-to-one correspondence to the diabatic states discussed above.
Panel (a) shows the time-dependent populations of the excited states.
Initially, all population is either in $S_2$ or $S_3$ (in the figure, their populations is combined as $S_{2,3}$), but the population is quickly transferred to the $S_1$ and subsequently to the triplet states.
In panel~(b), we show the net surface hops between the excited states.
As can be seen, most trajectories take a $S_{2,3}\rightarrow S_1\rightarrow T_{2,3}\rightarrow T_1$ path, as anticipated from the PESs shown in Figure~\ref{fig:liic}.
Hence, $S_{2,3}\rightarrow S_1\rightarrow T_{2,3}\rightarrow T_1$ would be a good kinetic model to fit the populations shown in Figure~\ref{fig:pop_const} (a).
However, in order to get time constants consistent with Ref.~\onlinecite{Mai2016JPCL}, we use the same kinetic model as in this reference; the model is shown in panel (c).
Using this kinetic model, we obtain three time constants describing the population transfer in the trajectories: (i) a time constant of 250~fs for $S_{2,3}\rightarrow S_1$ IC, (ii) a constant of 1060~fs for $S_{2,3}\rightarrow T_{2,3}$, and (iii) a time constant of 325~fs for $S_1\rightarrow T_1$.
The corresponding time constants from Ref.~\onlinecite{Mai2016JPCL} are (i) 59~fs, (ii) 250~fs, and (iii) 540~fs.

There are some differences to the population flow as reported in Ref.~\onlinecite{Mai2016JPCL}.
First, with MS-CASPT2, more trajectories hop through the side channels $S_1\rightarrow T_1$ and $S_2\rightarrow T_{2,3}$, while with RI-ADC(2) most trajectories hop according to the main channel $S_{2,3}\rightarrow S_1\rightarrow T_{2,3}\rightarrow T_1$.
Second, with MS-CASPT2, there was a tiny contribution of ground state relaxation from the $S_1$; with RI-ADC(2) ground state relaxation did not take place.
Third, with MS-CASPT2, singlet IC is notably faster than with RI-ADC(2) (59~fs versus 250~fs); this might be due to the different preferences of Paths I and II by the two methods. 
In particular, RI-ADC(2) has similar barriers for both paths, whereas Path II at MS-CASPT2 level is nearly barrierless.
Furthermore, with RI-ADC(2), Path I involves pyramidalization in $S_2$, which takes some time, while with MS-CASPT2, the $S_1$ can be reached quickly before the slow pyramidalization starts.
Fourth, the ISC time constants are different for the two methods.
This is partially because the ISC time constants are coupled to the IC time constant (because ISC is a follow-up reaction).
As can be seen, the ratio between the $S_{2,3}\rightarrow S_1$ and $S_{2,3}\rightarrow T_{2,3}$ time constants is about 1:4 for both electronic structure methods.
Hence, after 1000~fs, both methods provide very similar populations of the excited states: 65\% $T_1$, 20\% $T_{2,3}$, and 15\% $S_1$ for MS-CASPT2; 70\% $T_1$, 20\% $T_{2,3}$, and 10\% $S_1$ for RI-ADC(2).
Thus, it appears that, even though the time constants themselves are different, the interplay of the time constants lead to similar overall results.

\begin{figure}
  \centering
  \includegraphics[scale=1]{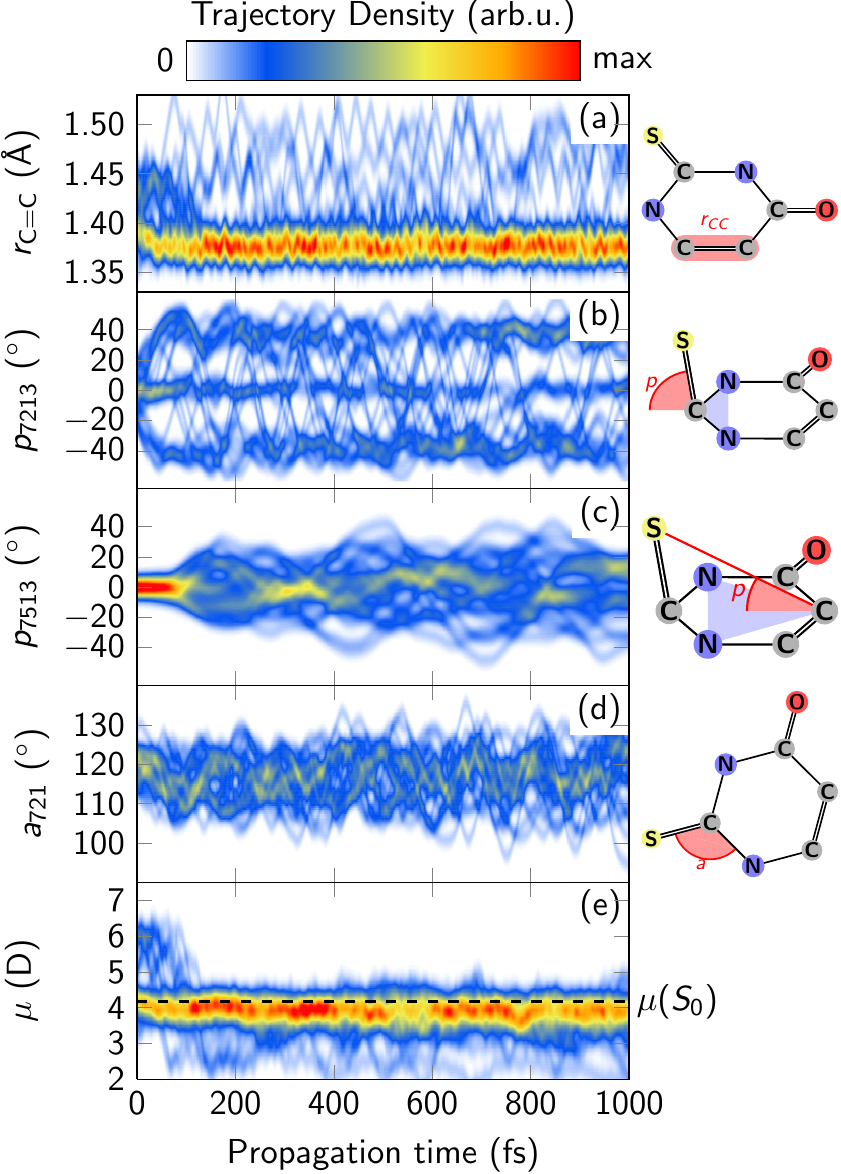}
  \caption{Time-dependent distribution of internal coordinates of 2TU with RI-ADC(2)/def2-SVP. The shown internal coordinate is sketched on the right and highlighted in red.}
  \label{fig:molecular_motion}
\end{figure}

In Figure~\ref{fig:molecular_motion}, we show an analysis of the molecular motion of 2TU during the simulation, similar to that in Ref.~\onlinecite{Mai2016JPCL}.
In panel (a), we plot the distribution of the C=C bond length over time, which is an indicator mode to distinguish the $^1\pi_S\pi^*_6$ ($S_2$) state from the other states ($^1\pi_S\pi^*_2$, $^1n_S\pi^*_2$, $^3\pi_S\pi^*_2$).
The former state has a bond length of about 1.44~\AA, whereas the other states have about 1.37~\AA.
The panel shows clearly that only few trajectories show the long bond typical for the $^1\pi_S\pi^*_6$ state, in contrast to the MS-CASPT2 dynamics, where initially all trajectories exhibit the long bond length.

In panel (b), we plot the pyramidalization angle of the thiocarbonyl group, one of the most important modes in the excited-state dynamics of 2TU.
The plot shows that pyramidalization starts immediately after excitation; in MS-CASPT2, pyramidalization only starts after about 50~fs, when the trajectories change from the planar $^1\pi_S\pi^*_6$ state to the pyramidalized $^1n_S\pi^*_2$ state.
The plots in panels (a) and (b) unambiguously show that RI-ADC(2) prefers Path I (see Figure~\ref{fig:liic}) over Path II, in opposition to MS-CASPT2.

Panel (c) shows another pyramidalization angle, which measures how much the sulfur atom is displaced from the ring plane while ignoring the motion of the thiocarbonyl carbon atom.
This panel is in good agreement with the MS-CASPT2 dynamics, showing that the long-term motion of the molecule in the $S_1$ and $T_1$ states is correctly captured.
The same can be said for panel (d), which shows the bond angle which controls ISC by tuning the energy gap between the $^1n_S\pi^*_2$ and $^3\pi_S\pi^*_2$ states, which are primarily involved in the ISC process.

Finally, in panel (e) we plot the temporal evolution of the permanent dipole moment of 2TU.
Like panels (a) and (b), this panel also shows that RI-ADC(2) favors Path I, as the more polar $^1\pi_S\pi^*_6$ state (at about 6~D) is not strongly populated at early times.
Moreover, the panel shows that RI-ADC(2) provides very similar permanent dipole moments for the other excited states, which are all around 4.0~D and similar to the ground state dipole moment of 4.2~D.

As a brief remark, three ADC(2) trajectories followed a ring-opening pathway, with the bond between N$_3$ and C$_4$ breaking.
Additional calculations showed that the relevant barrier is about 4.0--4.2~eV above the ground state energy (with both ADC(2) and MS-CASPT2), and hence much higher than the $S_1$ and $T_1$ minima.
Consequently, the three trajectories all undergo ring opening from the $S_2$ state briefly after excitation.
The reason for the more frequent ring opening with ADC(2) compared to MS-CASPT2\cite{Mai2016JPCL} might be that ADC(2) predicts rather high vertical excitation energies and thus an increased initial energy, enough to overcome the ring opening barrier.
Still, as mentioned above, we neglected the ring-opening trajectories from the above analysis, as the $S_0-S_1$ energy gap becomes very small and ADC(2) is not expected to be reliable in this situation.

\subsection{Computational remarks}

After the discussion of the dynamics simulations results, here we would like to also mention the computational effort involved in the present calculations.
Each of the RI-ADC(2)/def2-SVP trajectories calculated was run on 2~cores of an Intel\textsuperscript{\textregistered} Xeon E5-2650-v3 CPU and the 2000~time steps were on average finished after 5~days, which is equivalent to 240 core hours per trajectory.
By comparison, the MS-CASPT2 trajectories reported in Ref.~\onlinecite{Mai2016JPCL} were run on 16~cores of an Intel\textsuperscript{\textregistered} Xeon E5-2650-v2 CPU, and completion of 2000~time steps took about 18000~core hours per trajectory.
Considering also that the MS-CASPT2 trajectories included one state less (the $S_3$) and that due to convergence problems a sizable fraction of the MS-CASPT2 trajectories had to be neglected, it appears that RI-ADC(2) was over 100 times more CPU-time-efficient than MS-CASPT2 for the chosen molecule.
Due to the scaling behaviors of the two methods, it can be expected that for larger systems RI-ADC(2) will be even more favorable.
Furthermore, RI-ADC(2) offers advantages due to its conceptual simplicity owing to the fact the no active space has to be selected; this not only makes the computations more user friendly, but also offers more stable convergence behavior along a trajectory and allows studying systems where many relevant orbitals are involved.


\section{Conclusions}

In this contribution, a new implementation is reported which allows performing surface hopping simulations of ISC processes in the \textsc{Sharc} dyanmics package in combination with the ab-initio RI-ADC(2) electronic structure method---a computationally efficient, reliable, and easy-to-use method.
We expect that this new implementation will provide a powerful tool for simulating ISC between one-electron excited states in a large variety of organic molecules.
For completeness, it should be pointed out that RI-ADC(2) is not expected to work in some special cases involving two-electron excited states,\cite{Starcke2006} transition metal complexes,\cite{Plasser2015JPCA} or strongly distorted geometries.
Furthermore, RI-ADC(2) cannot correctly describe conical intersections between the ground state ($S_0$) and the excited states.
In such cases the user is advised to use multi-reference methods, e.g., as those already available in \textsc{Sharc}.\cite{Mai2014JCP_SO2,Mai2016JPCL}

The 2-thiouracil molecule was chosen to demonstrate the capabilities of the new implementation.
First, the vertical excitations were compared to the more sophisticated ADC(3) and MS-CASPT2 methods.
While this comparison showed some differences between the methods, RI-ADC(2) provided an excellent description of all the states relevant for the dynamics of this molecule.
Furthermore, the SOC terms were compared between the RI-ADC(2) and MS-CASPT2 methods, showing satisfactory agreement.
Finally, RI-ADC(2) dynamics simulations were performed and compared to previously reported MS-CASPT2 results.\cite{Mai2016JPCL}
Slightly different mechanistic details were obtained, as $S_2\rightarrow S_1$ IC is accompanied at the RI-ADC(2) level by pyramidalization in the $S_2$ state, whereas MS-CASPT2 predicts that pyramidalization is initiated in the $S_1$ state.
Nevertheless, a good agreement between both methods is found in terms of the main deactivation mechanisms, i.e., that IC to the $S_1$ state is followed by ISC, which leads to the population of the $T_1$ state in less than 1~ps.


\section*{Supplementary Material}

Coordinates of the geometry and active space orbitals used for the results presented in section IV A.
Natural transition orbitals of 2TU for the RI-ADC(2), ADC(3), and MS-CASPT2 calculations in section IV A.


\section*{Acknowledgments}

S. M., F. P., and L. G. thank the Austrian Science Fund (FWF) for funding through project P25827, the University of Vienna, and the Vienna Scientific Cluster (VSC3) for generous allocation of computing time.

A.K. and M.P. acknowledge generous support by Prof.\ J. Gauss during their stay in Mainz. 
A.K. also thanks Prof.\ M. Kallay for helpful discussions about ADC(2) transition densities. 

The authors also acknowledge the COST action CM1305 (ECOSTBio).


\appendix
\section{Spin-free formulation of triplet transition densities}
\label{app:ttd}

The implementation in \textsc{Turbomole} allows for the computation of all spin-orbit matrix elements in both the spin-orbital basis and in a spin-free formulation. As the spin-orbital formulation does not offer explicit control of the spin state, its use for open-shell ground states is not recommended. For closed-shell ground states, a more efficient spin-free formulation for both singlet and triplet excited states is available that has been tested against the spin-orbital code. The spin-free cluster operator is defined as
\begin{align}
 \hat T_2 = \frac{1}{2}\sum_{ijab} t_{iajb} \hat E_{ai} \hat E_{bj}
\end{align}
where
\begin{align}
 t_{iajb} = -\frac{g_{iajb}}{\epsilon_a+\epsilon_b-\epsilon_i-\epsilon_j}
\end{align}
are the first-order amplitudes from MP2 theory. These are computed from two-electron repulsion integrals $g_{iajb}$ and canonical orbital energies $\epsilon_p$.

The singlet excited state $I$ is associated with the excitation operator $\hat R^{I} = \hat R^{I}_1+\hat R^{I}_2$, with single and double excitations
\begin{align}
 \hat R^{I}_1 &= \sum_{ia} R^{I}_{ia} \hat E_{ai}\\ 
 \hat R^{I}_2 &= \frac{1}{2}\sum_{ijab} R^{I}_{iajb} \hat E_{ai} \hat E_{bj}\,.
\end{align}
For the formulation of triplet excited states, we introduce the triplet excitation operator
$\hat T_{ai} = 2 \hat E^z_{ai} = \exop{a}{i}-\exop{\bar a}{\bar i}$
and define the single and double excitations as
\begin{align}
 {}^3\!\hat R^{I}_1 &= \sum_{ia} {}^3\!R^{I}_{ia} \hat T_{ai} \\
 {}^3\!\hat R^{I}_2 &= \sum_{ijab} {}^3\!R^{I}_{iajb} \hat T_{ai} \hat E_{bj}\\
                    &= \frac{1}{4}\sum_{ijab} {}^{(+)}\!R^{I}_{iajb} \left(\hat T_{ai} \hat E_{bj}+\hat E_{ai} \hat T_{bj}\right)\\
                    &+\frac{1}{2}\sum_{ijab} {}^{(-)}\!R^{I}_{iajb} \left(\hat T_{ai} \hat E_{bj}-\hat E_{ai} \hat T_{bj}\right)
\end{align}
The symmetrized formulation has been given for comparison with the triplet basis introduced in Refs. \cite{Hald:CPL328-291,Haettig:JCP116-5401}. The ADC(2) equations using the singlet and triplet adapted excitation operators can be easily derived from the more general CC2 equations given in Refs.  \cite{Hald:CPL328-291,Haettig:JCP113-5154,Haettig:JCP116-5401,Haettig:JCP117-6939}.

We only want to explicitly quote the spin-transition densities from ground to excited states, $D_{pq}^{0J,z}=\bra{\Psi_0}\hat{E}_{pq}^{z}\ket{\Psi_J}$ and between excited states $D_{pq}^{IJ,z}=\bra{\Psi_I}\hat{E}_{pq}^{z}\ket{\Psi_J}$. The ground-to-excited-state transition density reads
\begin{align}
D_{ij}^{0J,z} &= -\frac{1}{\sqrt{2}} \sum_{cdl}  t^\dagger_{cjdl} {}^3\!R^{J}_{icld}\\
D_{ia}^{0J,z} &= \frac{1}{\sqrt{2}}  {}^3\!R^J_{ia}\\
D_{ai}^{0J,z} &= -\frac{1}{\sqrt{2}}\sum_{ck}  t^\dagger_{ciak}  {}^3\!R^{J}_{kc}\\
D_{ab}^{0J,z} &= \frac{1}{\sqrt{2}}\sum_{dkl} t^\dagger_{akdl} {}^3\!R^{J}_{kbld}
\end{align}
and the transition density between excited states is given by
\begin{align}
D_{ij}^{IJ,z} &= \frac{1}{2}\left[ -\sum_{c}  R^{I\,\dagger}_{cj} {}^3\!R^{J}_{ic} - \sum_{cdl}  R^{I\,\dagger}_{cjdl} {}^3\!R^{J}_{icld}\right]\\
D_{ia}^{IJ,z} &= \frac{1}{2}\left[ \sum_{dl}R^{I\,\dagger}_{dl}{}^3\!R^{J}_{iald}  - \sum_{cdkl} \left[2 R^{I\,\dagger}_{ckdl} - R^{I\,\dagger}_{dkcl}\right] t_{kald} {}^3\!R^{J}_{ic}\right.\nonumber\\
              &  \left.- \sum_{cdkl} \left[2 R^{I\,\dagger}_{ckdl} - R^{I\,\dagger}_{dkcl}\right] t_{icld} {}^3\!R^{J}_{ka}\right]\\
D_{ai}^{IJ,z} &= \frac{1}{2}\left[ -\sum_{dl}R^{I\,\dagger}_{dial}{}^3\!R^{J}_{ld}  - \sum_{cdkl} R^{I\,\dagger}_{ci} t^\dagger_{akdl} {}^3\!R^{J}_{kcld} \right.\nonumber\\
              &  \left.- \sum_{cdkl}R^{I\,\dagger}_{ak} t^{\dagger}_{ckdl}  {}^3\!R^{J}_{icld} \right]\\
D_{ab}^{IJ,z} &= \frac{1}{2}\left[\sum_{k} R^{I\,\dagger}_{ak} {}^3\!R^{J}_{kb}+ \sum_{dkl} R^{I\,\dagger}_{akdl} {}^3\!R^{J}_{kbld}\right]
\end{align}
In the actual implementation, all four-index quantities are on-the-fly recomputed from three-index quantities using the resolution-of-the-identity trick as described in Refs. \cite{Haettig:JCP116-5401,Haettig:JCP117-6939}.



%

\end{document}